\documentclass[aps,pre,twocolumn,superscriptaddress,showkeys]{revtex4-2}

\pdfoutput=1

\usepackage[T1]{fontenc}
\usepackage{amsmath,amssymb}
\usepackage{graphicx}
\usepackage{dcolumn}
\usepackage{bm}
\usepackage{float}
\usepackage{color}
\usepackage{amsmath}
\definecolor{rojo}{rgb}{1,0,0}
\definecolor{verde}{rgb}{0,0.8,0.2}
\definecolor{azul}{rgb}{0,0,1}
\definecolor{rosa}{cmyk}{0,1,0,0}
\usepackage{appendix}
\usepackage{amsmath}
\usepackage{amsthm}
\usepackage{amsfonts}
\usepackage{amssymb}
\usepackage{color}
\usepackage{epsfig}
\usepackage{hyperref}
\hypersetup{colorlinks=true, citecolor=blue}
\usepackage{soul}
\usepackage{appendix}
\usepackage{subcaption}

\newcommand{\jp}[1]{{\color{black}#1}}

\begin{document}
\preprint{APS/123-QED}

\title{Additional jamming transition in 2D bidisperse granular packings}

\author{Juan C. Petit}
\affiliation{Institute of Materials Physics in Space, German Aerospace Center (DLR), 51170 K\"oln, Germany}
\author{Matthias Sperl}
\affiliation{Institute of Materials Physics in Space, German Aerospace Center (DLR), 51170 K\"oln, Germany}
\affiliation{Institut f\"{u}r Theoretische Physik, Universit\"{a}t zu K\"{o}ln, 50937 K\"{o}ln, Germany}%

\date{\today}
\begin{abstract}
We present a jamming diagram for 2D bidisperse granular systems, capturing two distinct jamming transitions. 
The first occurs as large particles form a jammed structure, while the second, emerging at a critical small-particle 
concentration,  $X_{\mathrm{S}}^{*} \approx 0.21$, and size ratio, $\delta^{*} \approx 0.25$, involves  
small particles jamming into the voids of the existing large-particle structure upon further compression.
Below this threshold, small particles fill voids within the large-particle network, increasing packing density. 
Beyond this point, excess small particles disrupt efficient packing, resulting in looser structures.
\jp{These results, consistent with previous 3D studies, demonstrate that the second transition occurs at a 
well-defined point in the $(X_{\mathrm{S}}, \delta)$ plane, independent of dimensionality, likely driven by 
the geometric saturation of available space around particles, void closure, and structural arrangement.}
\end{abstract}

\maketitle

The jamming transition occurs when granular materials shift from a fluid-like 
state, where particles can move freely, to a solid-like state, where most particles 
become immobilized. This phenomenon arises at a critical density, $\phi_{J}$, when 
increasing compression or particle packing density restricts their motion, effectively 
freezing the system into a jammed configuration.
This transition has been extensively studied in both two- and 
three-dimensional monodisperse and bidisperse packings \cite{donev2004jamming, meyer2010jamming,
koeze2016mapping, o2003jamming, majmudar2007jamming, van2009jamming, 
behringer2018physics, prasad2017subjamming, odagaki2002random, atkinson2014existence, biazzo2009theory, 
hopkins2011phase, kumar2016tuning, petit2020additional, hara2021phase}. 
In monodisperse sphere packings, the jamming transition occurs at $\phi_{J} \approx 0.64$ 
\cite{donev2004jamming, o2003jamming, van2009jamming, behringer2018physics}. In contrast, 
bidisperse packings exhibit a broader range of $\phi_{J}$ values, increasing with 
lower size ratios $\delta$ and lower concentrations of small particles $X_{\mathrm{S}}$ 
\cite{prasad2017subjamming, biazzo2009theory, hopkins2011phase, kumar2016tuning, hara2021phase, furnas1931grading}.
Studies reveal that in asymmetric bidisperse packings, the system transitions from 
a small-sphere-rich to a small-sphere-poor structure. This transition is caused by 
an abrupt drop in the number of small particles contributing to the jammed structure 
at a specific $X_{\mathrm{S}}$, leaving the remaining small particles without 
contacts \cite{prasad2017subjamming, hara2021phase}. Recent work \cite{petit2020additional} 
identified an additional jamming transition line in 3D bidisperse systems, arising 
from the jamming of small particles that were previously without contacts. An emerging 
point is observed at $X_{\mathrm{S}}^{*}(\delta^{*}) \approx 0.21$ with $\delta^{*} \approx 0.22$, 
below which two distinct jamming transitions occur: one dominated by large particles at 
lower $\phi$, followed by a second, discontinuous jamming of small particles 
at higher $\phi$. This second transition not only exhibits unique mechanical 
properties, as demonstrated in Refs.~\cite{petit2022bulk, petit2023structural}, 
but also enriches the jamming diagram for 3D bidisperse packings.

\begin{figure*}[t]  
\centering
\begin{subfigure}[t]{0.57\textwidth}
    \centering
    \includegraphics[width=\linewidth]{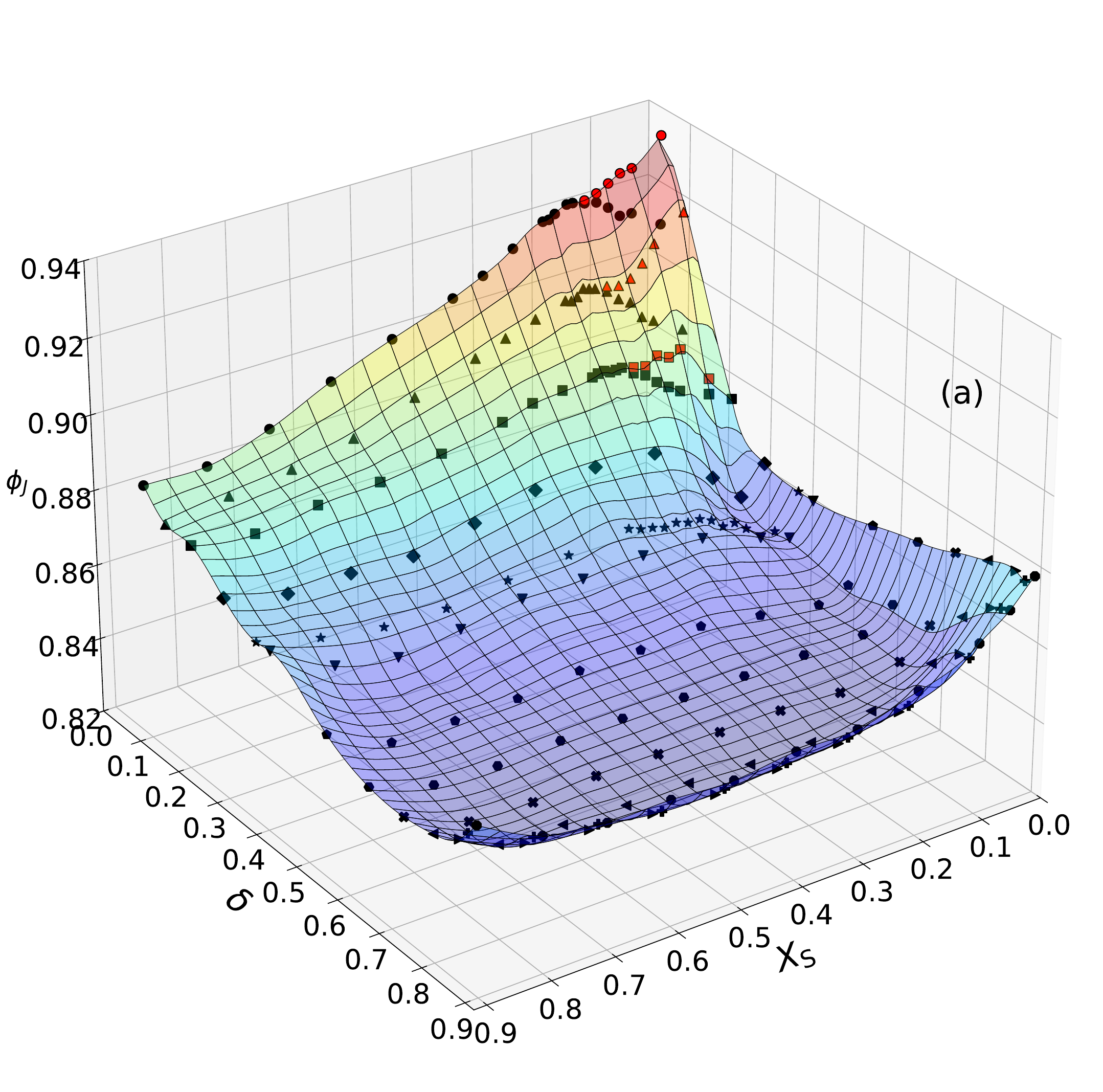}
    \label{fig:img1}
\end{subfigure}
\hfill
\begin{subfigure}[t]{0.42\textwidth}
    \centering
    \includegraphics[width=\linewidth]{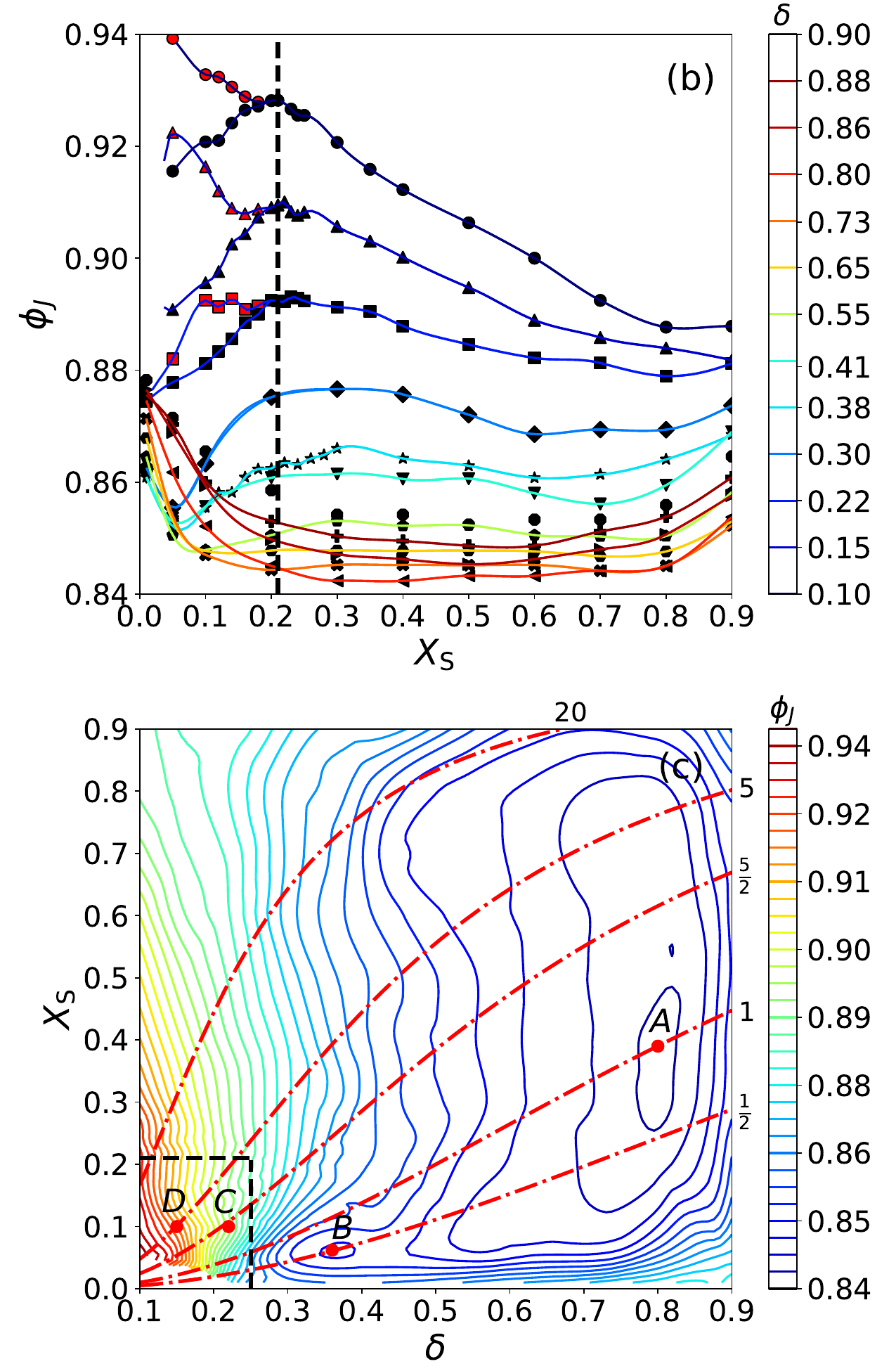}
    \label{fig:img2}
\end{subfigure}
\caption{Surface plot of the jamming density $\phi_J$ as a function 
of size ratio $\delta$ and small particle concentration $X_{\mathrm{S}}$, 
with the color map representing $\phi_J$. \jp{Black symbols (first transition) and 
red symbols (second transition) show simulation results averaged over three 
realizations, with standard deviations below $10^{-2}$ (not shown)}. These 
data points guide the surface fit and delineate the two jamming regimes. 
(b) Projection of $\phi_J$-$X_{\mathrm{S}}$ for 
specific $\delta$ values from panel (a). (c) Contour plot of $\phi_J$ in 
the $(\delta, X_{\mathrm{S}})$ plane. Red dash-dotted lines show 
$X_{\mathrm{S}}$-$\delta$ relations for various $N_{\mathrm{S}}/N_{\mathrm{L}}$ 
ratios. Red dots (A-D) mark configurations in Fig.~\ref{fig3}. Black dashed 
lines indicate the emerging point at $X_{\mathrm{S}}^{*} \approx 0.21$ and 
$\delta^{*} \approx 0.25$, below which the second jamming transition occurs.}
\label{fig1}
\end{figure*}

While the behavior of 3D packings reveals intricate jamming phenomena, 2D 
monodisperse and bidisperse systems are widely used to simplify and deepen 
our understanding. In polycrystalline monodisperse disk packings, 
the jamming transition occurs around $\phi_{J} \approx 0.88$ 
\cite{donev2004jamming, meyer2010jamming}, whereas in disordered configurations, 
it drops to approximately $\phi_{J} \sim 0.81$ \cite{meyer2010jamming, odagaki2002random, 
atkinson2014existence}. In binary systems with a size ratio of $\delta = 0.71$
and equal particle concentration (50:50), jamming occurs at $\phi_{J} \approx 0.84$ 
\cite{majmudar2007jamming, o2003jamming}. Systematic studies of the jamming transition 
have revealed a complex jamming diagram across the range $\delta, X_{\mathrm{S}} \in [0,1]$, 
see Ref.~\cite{odagaki2002random}, with a more detailed mapping provided in Ref.~\cite{koeze2016mapping}. 
These works demonstrate a maximum $\phi_{J}$ at low $\delta$ and high $X_{\mathrm{S}}$. 
Notably, Ref.~\cite{koeze2016mapping} highlights that the maximum $\phi_{J}$ is accompanied 
by a high number of small particles acting as rattlers, trapped in gaps between large particles 
and contributing minimally to the jammed structure.
This behavior points to a decoupling in the jamming process of 
small and large particles, suggesting the presence of an additional transition 
line within the packing. In this work, we explicitly demonstrate that a second 
jamming transition can be rigorously identified, similar to that observed in 3D 
bidisperse packings \cite{petit2020additional}. This additional transition line is 
identified for $X_{\mathrm{S}} < X^{*}_{\mathrm{S}}(\delta^{*}) \approx 0.21$ and 
$\delta < \delta^{*} \approx 0.25$, distinguishing a jammed structure formed 
solely by large particles at low $\phi$ from one that incorporates both small 
and large particles at higher $\phi$. The transition diagram for this case is 
illustrated in Fig.~\ref{fig1} and will be further discussed below.

2D bidisperse packings are simulated using the MercuryDPM software 
\cite{weinhart2020fast}, which employs the discrete element method (DEM). 
Packings consist of $N = 5000$ particles, with a 
number of large, $N_{\mathrm L}$, and small, $N_{\mathrm S}$, particles and radius 
$r_{\mathrm L}$ and $r_{\mathrm S}$, respectively. The size ratio $\delta = r_{\mathrm 
S}/r_{\mathrm L}$, concentration of small particles, $X_{\mathrm 
S} = N_{\mathrm S} \delta^{2} / (N_{\mathrm L} + N_{\mathrm S} 
\delta^{2})$, and the overall packing fraction $\phi$ characterize the bidispersity
of the system.  An isotropic deformation, similar to that in Ref.~\cite{petit2020additional}, 
is applied here using the Hertzian spring dashpot contact model under the assumption of zero friction.
Each bidisperse packing, defined by parameters ($\delta, X_{\mathrm{S}}$), is generated 
and subsequently compressed and decompressed following a standardized protocol, 
as described in Sec.~I of Supplemental Material \cite{SupplementalMaterial}. 
\jp{The jamming density, $\phi_J$, at the first (second) transition 
is defined as the packing fraction at which a sudden increase is observed in the fraction 
of large (small) particles contributing to the jammed structure}. These fractions are defined 
as $n_{\mathrm{L}}(\phi) = N^{c}_{\mathrm{L}} / N$ and $n_{\mathrm{S}}(\phi) = N^{c}_{\mathrm{S}} / N$, 
where  $N^{c}_{\mathrm{L,S}}$ represents the number of large and small particles in contact 
(see Sec.~II of \cite{SupplementalMaterial}). Fig.~\ref{fig1} (a) presents a surface
jamming diagram $(\phi_J, \delta, X_{\mathrm{S}})$, constructed via cubic interpolation of 
the measured $\phi_J$ values. The surface shows minima in $\phi_J$ at intermediate $\delta$ and a strong 
enhancement at low $\delta$ and $X_{\mathrm{S}}$, with the color map emphasizing these variations. Different 
colored symbols represent simulation data and are overlaid to highlight the transitions: 
black symbols indicate the first jamming transition, while red symbols denote the second. These transitions
reflect distinct size-disparity effects, with further analysis given in 
Fig.~\ref{fig1} (b)-(c).

\begin{figure}[t]
\centering \includegraphics[scale=0.32]{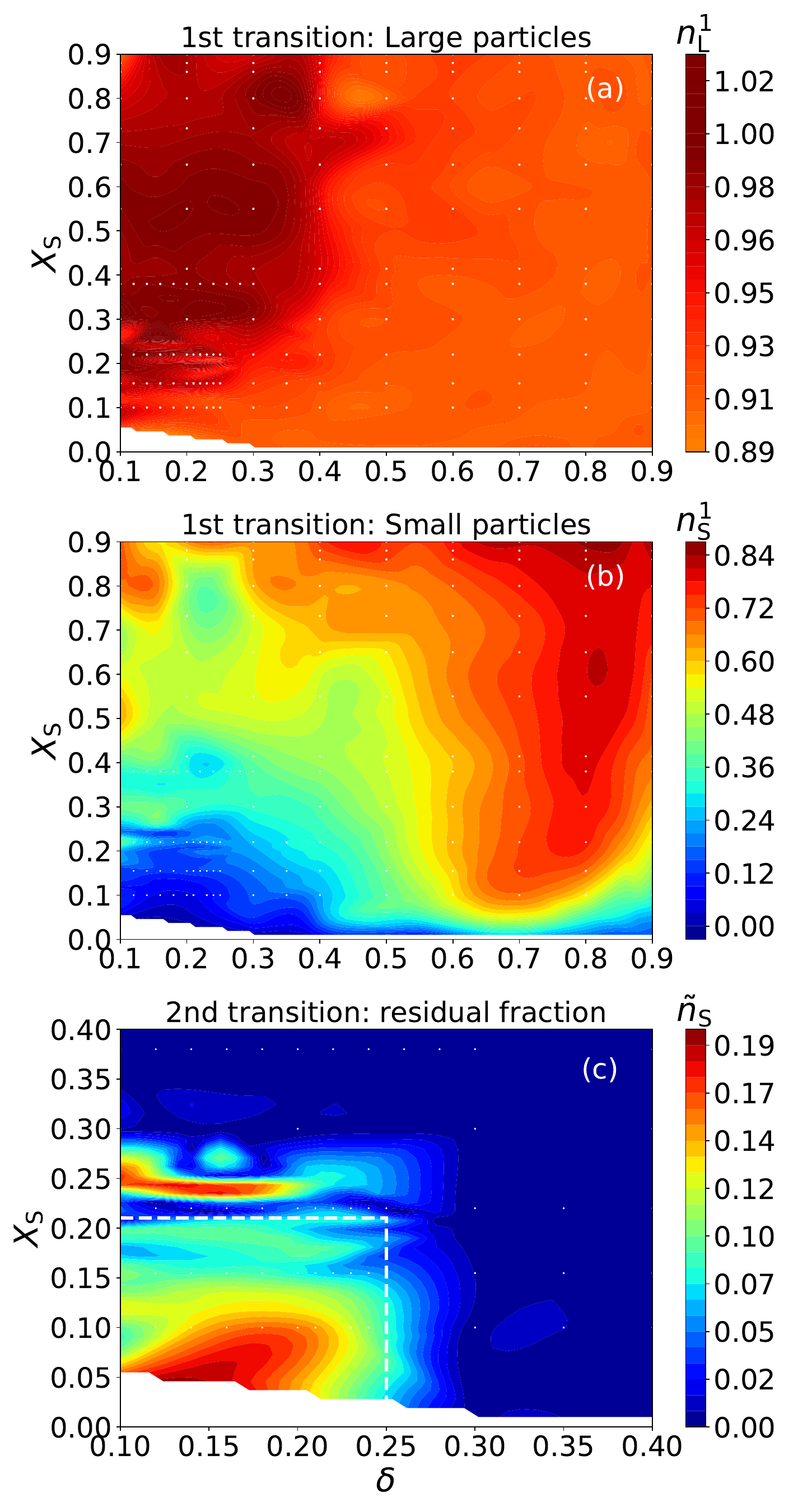}
\caption{Contribution of the non-rattler particle fraction ($Z \geq 4$) 
to the jamming transitions as a function of 
$\delta$ and $X_{\mathrm{S}}$. (a)-(b) fraction of large, $n^{1}_{\mathrm{L}}$, and small 
particles, $n^{1}_{\mathrm{S}}$, at the first jamming transition. (c) residual fraction of
small particles, $\tilde{n}_{\mathrm{S}} = n^{2}_{\mathrm{S}} - n^{1}_{\mathrm{S}}$, at the 
second transition. $n^{2}_{\mathrm{S}}$ at the second transition 
is included in the Supplemental material \cite{SupplementalMaterial}. The 
emerging point is found at $X^{*}_{\mathrm{S}}(\delta^{*}) \approx 0.21$ and $\delta^{*} \approx 0.25$, see 
white dashed lines.}
\label{fig2}
\end{figure}

\begin{figure}[t]
\centering \includegraphics[scale=0.9]{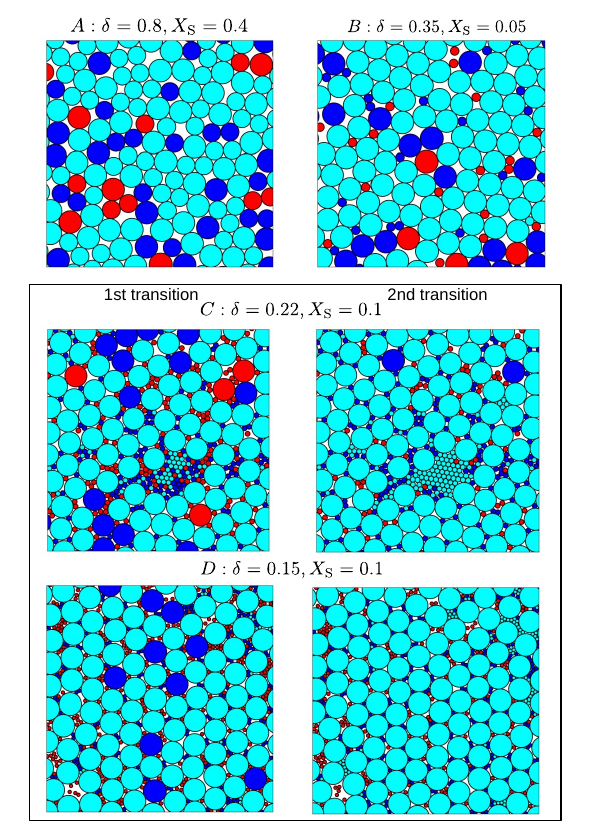}
\caption{\jp{Particle configurations at the first and second jamming transitions for various binary packings 
shown in Fig.~\ref{fig1}(c). Cyan, blue, and red disks indicate particles with 
$Z \geq 4$, $Z = 3$, and $Z < 3$, respectively, marking rigid, marginally 
stable, and rattler particles. Color changes at the second transition reveal how large and small 
particles contribute to the evolving jammed backbone. These configurations offer a close-up 
view of representative regions within the overall system.}}
\label{fig3}
\end{figure}

Fig.~\ref{fig1} (b) presents the projected $\phi_J$-$X_{\mathrm{S}}$ relationship (solid lines) 
with corresponding simulation results (colored symbols) from Fig.~\ref{fig1}(a).
For $\delta \geq 0.65$, $\phi_{J}$ shows a slightly 
flat behavior for $X_{\mathrm S} \in [0.1, 0.8]$ and a
sudden increase at extreme values. For $\delta < 0.65$, $\phi_{J}$ increases exhibiting 
a maximum value at specific $X_{\mathrm S}$, becoming prominent for lower $\delta$.
For $\delta \leq 0.22$, two distinct jamming transitions 
emerge from a common critical composition at $X_{\mathrm{S}}^{*} \approx 0.21$, which we 
define as the emerging point. This bifurcation marks the onset of fundamentally different 
jamming regimes, separating regions where the system exhibits 
either a single ($X_{\mathrm{S}} > X_{\mathrm{S}}^{*}$) or two jamming transitions
($X_{\mathrm{S}} < X_{\mathrm{S}}^{*}$). Typically, at low $\phi$, the large particles 
jam first, defining the first jamming transition line (black symbols), while most small 
particles remain unjammed and do not yet contribute to the jammed structure. 
Upon further compression, the small particles become incorporated into the jammed network, 
leading to a second jamming transition (red symbols). This two-step process is most 
pronounced at low $\delta$ 
and low $X_{\mathrm{S}}$, as illustrated in Figs.~\ref{fig1} (a)-(b). The shape 
of this second line depends on $\delta$. For $\delta = 0.22$, a plateau in $\phi_J$ is observed. 
This plateau arises because small particles cannot fit into a triangular lattice of large particles
after compression (see Fig.~\ref{fig3} C), resulting in a constant $\phi_J$ over a range of 
$X_{\mathrm{S}}$ values. The decrease in $\phi_J$ at lower $X_{\mathrm{S}}$ indicates that 
the small particle count is insufficient to significantly enhance the overall density.
For $\delta = 0.15$, small particles can now fit into the triangular lattice of 
large particles (see Fig.~\ref{fig3} D), leading to an enhancement in $\phi_J$, 
see Refs.~\cite{petit2023structural, kumar2016tuning}. A distinct scenario emerges for 
$\delta = 0.1$, where the small particles are so small that a substantial number are needed 
to fill the voids within the triangular lattice formed by the large particles. The 
overcompression required to jam the small particles alongside the large ones, resulting 
in the second transition, leads to a noticeable increase in the average particle overlap 
compared to the first jamming transition. This behavior is detailed in Sec.~III of 
Ref.~\cite{SupplementalMaterial}, where the average overlap along the second transition 
is shown to remain below $1\%$ of the large particle diameter, decaying nearly to zero 
at the first transition. These overlap values provide strong evidence supporting the 
validity of the second jamming transition even after further compression.

\jp{Although the second jamming transition in 2D stems from the same physical mechanism 
as in 3D—jamming of small particles under continued compression—the Furnas model fails to 
capture either transition in 2D. While it successfully reproduces both transitions in 3D 
systems \cite{petit2020additional, petit2022bulk}, it breaks down in 2D, as shown in Sec.~IV 
of Ref.~\cite{SupplementalMaterial}. This is due to geometric constraints: in 3D, smaller 
particles can hierarchically fill volumetric voids (e.g., tetrahedral or octahedral), making 
the model’s assumptions viable. In 2D, the flat voids between disks are too constrained to 
fit smaller particles without disrupting the packing, causing the model to overestimate 
density and miss the transitions. This challenges the Furnas model’s universality and 
calls for frameworks that explicitly account for dimensionality and structural heterogeneity.}

The first and second jamming transition lines are identified by analyzing the fractions of 
small, $n_{\mathrm{S}}^{X}$, and large, $n_{\mathrm{L}}^{X}$, particles participating in 
the jammed structure, where the superscript $X \in \{1, 2\}$ denotes the corresponding 
jamming stage. \jp{Particles are considered part of the globally jammed backbone if they 
satisfy the criterion $Z \geq 4$, consistent with the isostatic condition for mechanical 
stability in 2D frictionless packings. Although $Z \geq 3$ ensures local stability 
\cite{Silbert2006Structural}, the stricter $Z \geq 4$ threshold isolates the subset of 
particles that actively sustain the global jammed network and define the structural 
backbone of the jammed state.}
Fig.~\ref{fig2} shows $n_{\mathrm{S}}$ and $n_{\mathrm{L}}$ 
at the respective jamming transitions in the full range of $\delta$ and $X_{\mathrm{S}}$ values.
For $\delta \geq 0.6$, where the flat behavior of $\phi_J$ is observed for $X_{\mathrm{S}}$ 
in Fig.~\ref{fig1} (b), $n^{1}_{\mathrm{L}} \geq 90\%$ and 
$n^{1}_{\mathrm{S}} \geq 70\%$, which means that most of the large and small 
particles form the jammed structure, see Fig.~\ref{fig2} (a)-(b). 
For $0.4 < \delta < 0.6$, $n^{1}_{\mathrm{L}}$ remains constant, 
while $n^{1}_{\mathrm{S}}$ drops below $60\%$. Despite this reduction, large and small particles
still contribute simultaneously to the jammed structure. 
For $\delta \leq 0.4$ and $X_{\mathrm{S}} \lesssim 0.25$, $n^{1}_{\mathrm{S}}$
is low, while $n^{1}_{\mathrm{L}}$ remains high, see Fig.~\ref{fig2} (a)-(b). This implies 
that only large particles contribute to the jammed structure, while most small particles remain out of contact, 
see Fig.~2 (a,d) and (b,e) in Sec.~II of \cite{SupplementalMaterial}. This behavior 
defines the first jamming transition. With further compression, the fraction of small particles 
exhibits a sudden jump at higher $\phi$, and after a significant incorporation into the already 
jammed structure of large particles defines the second jamming transition, see Fig.~\ref{fig1} (b)
and Fig.~2 (c,f) in \cite{SupplementalMaterial}.

The fraction of small particles at the second transition, $n^{2}_{\mathrm{S}}$,  
is similar to Fig.~\ref{fig2} (b) except for a subtle change at low $\delta$ 
and low $X_{\mathrm{S}}$, see Sec.~V in \citep{SupplementalMaterial}.
To highlight this region, we quantify 
the difference $\tilde{n}_{\mathrm{S}} = n^{2}_{\mathrm{S}} - n^{1}_{\mathrm{S}}$, which 
captures the change in the fraction of small particles between the first and second transitions. 
For instance, $\tilde{n}_{\mathrm{S}} = 0$, signifies that the fraction of small particles 
remains unchanged at the second transition, confirming that they become jammed together 
with the large particles during the first transition.
Fig.~\ref{fig2} (c) shows the values of $\tilde{n}_{\mathrm{S}}$ at the second transition.
It illustrates a substantial increase in the number of small particles with $Z \geq 4$ contributing  
to the already jammed structure at low  $X_{\mathrm{S}}$ and low $\delta$. More than $10\%$ of 
small particles are jammed, 
responsible for the second jamming transition shown in Fig.~\ref{fig1} (a)-(b). 
The emerging point of the second transition is identified at 
$X^{*}_{\mathrm{S}}(\delta^{*}) \approx 0.21$ and $\delta^{*} \approx 0.25$ (see white dashed lines), 
as this marks the boundary where regions with a substantial number of small particles are 
surrounded by low values of 
$\tilde{n}_{\mathrm{S}}$. The smaller red region around 
$X_{\mathrm{S}} \approx 0.25$ (see Fig.~\ref{fig2} (c)) might be considered as packings 
experiencing the second transition as well; however, this is not the case, as both large
and small particles jam simultaneously in this region. This is just a continuous 
increase of the small particle number near the emerging point.

Fig.~\ref{fig1} (a) offers further insight that deepen our understanding
into the jamming diagram, complemented by Fig.~\ref{fig1} (c), which shows 
contour lines of constant $\phi_J$. Two minima are observed: (A) a global 
minimum at $\delta \approx 0.8$ and $X_{\mathrm{S}} \approx 0.4$, and (B) a local minimum 
at $\delta \approx 0.35$ and $X_{\mathrm{S}} \approx 0.05$, exhibiting a similar  
jamming density. Interestingly, the global minimum 
corresponds to a 50:50 particle mixture, as shown by the dash-dotted lines in Fig.~\ref{fig1} (c), 
which represent distinct ratios of small to large particles $(N_{\mathrm{S}} / N_{\mathrm{L}})$. 
Fig.~\ref{fig3} (A) provides a close-up view of the jammed particle configuration (cyan) at the
global minimum, highlighting locally stable large and small particles (blue) as well as rattlers (red).
In contrast, the local minimum at $\delta \approx 0.35$ corresponds to a packing where the number 
of small particles is roughly half that of the large ones. The particle configuration in 
Fig.~\ref{fig3} (B) reveals that a greater fraction of small particles remain disconnected from 
the jammed backbone. These minima correspond to loose packings characterized by the lowest 
jamming density  in bidisperse granular systems.

A jamming diagram surface similar to Fig.~\ref{fig1} (a) was previously reported in 
Ref.~\cite{koeze2016mapping}, identifying the same minima but at different $X_{\mathrm{S}}$ 
values. This discrepancy arises from differing definitions of small particle concentration: 
their work uses the small particle fraction, 
$f_{n} = N_{\mathrm{S}} / (N_{\mathrm{S}} + N_{\mathrm{L}})$, while we adopt the \jp{area 
fraction}, $X_{\mathrm{S}}$. These definitions are related via 
$X_{\mathrm{S}} = f_{n} \delta^{2} / (1 - f_{n} + f_{n} \delta^{2})$, allowing us to 
confirm consistency between our minima and theirs. A key distinction in our diagram is 
the presence of additional transition lines at low $\delta$ and $X_{\mathrm{S}}$, see 
red symbols in Fig.~\ref{fig1} (a)-(b), absent in Ref.~\cite{koeze2016mapping}. 
However, their study highlights an increase in small particle rattlers 
within the same $\delta$ and $X_{\mathrm{S}}$ range, indicating the potential for a
second transition. Black dashed lines in Fig.~\ref{fig1} (c) mark the emerging 
point at $(X_{\mathrm{S}}^{*}, \delta^{*}) \approx (0.21, 0.25)$, which define the 
boundary for consistent occurrence of the second jamming transition. Representative 
packings at $\delta = 0.22$ (C) and $\delta = 0.15$ (D) for $X_{\mathrm{S}} = 0.1$ 
demonstrate systems with approximately five and five-and-a-half times more 
small particles than large ones. At the second 
transition, $\delta = 0.22$ exhibits a higher $\phi_J$ compared to the first transition, 
due to the incorporation of rattler particles into the jammed structure upon compression, 
see Fig.~\ref{fig3} (C). In contrast, $\delta = 0.15$ represents a special 
size ratio where a single small particle can perfectly fill the void within a triangular 
lattice of large particles, satisfying 
$r_{\mathrm{S}} = \big[(2 - \sqrt{3})/\sqrt{3}\big] r_{\mathrm{L}}$. This configuration 
results in a denser packing and is observed in Fig.~\ref{fig3} (D), 
particularly at the second transition and occasionally at the first one. Note that 
the second transition can occur across a wide range of particle ratios 
$N_{\mathrm{S}} / N_{\mathrm{L}}$. For the commonly studied 50:50 mixture 
($N_{\mathrm{S}} / N_{\mathrm{L}} = 1$), the onset of the second 
transition is observed at $X_{\mathrm{S}} \leq 0.05$. Introducing particle ratios 
$N_{\mathrm{S}} / N_{\mathrm{L}}$ alongside the jamming diagram offers a clearer 
understanding of how large and small particles contribute to the jammed structure 
of the mixture.

In conclusion, we have constructed a detailed surface jamming diagram 
$(\phi_J, \delta, X_{\mathrm{S}})$ that presents both the first 
and second jamming transitions. The first transition is dominated by large 
particles, with small particles not contributing, while the second transition 
occurs as small particles jam into the already jammed structure of large particles. 
This second transition arises for $\delta \leq 0.25$ and $X_{\mathrm{S}} < 0.21$, a range 
where the size asymmetry is so pronounced that, beyond the typical configuration 
where a small particle fills the voids of a triangular arrangement of 
large particles, numerous small particles occupy interstitial spaces, 
leading to an increased jamming density with further compression.
Outside of this range, small particles do not fit into the void of 
large ones leading to a looser packing. The second transition line and the
emerging point values $\delta^{*} \approx 0.25$ and $X_{\mathrm{S}}^{*}(\delta^{*}) \approx 0.21$ 
align with 3D results obtained using the linear contact model \cite{petit2020additional}. 
\jp{This consistency indicates that the transition is independent of the contact model and 
might suggests that its emerging point is governed primarily by the packing geometry of the mixture, 
specifically by the saturation of local space, void closure, and the emergence of structural 
arrangement}.
While our focus here has been on the geometrical and structural aspects of the second 
jamming transition in two dimensions, a deeper understanding of its physical 
nature—particularly through critical scaling analysis—remains an important open 
question. Previous work in 3D systems \cite{petit2025pressure} shown distinct 
scaling behavior of the coordination number across the two transitions, while pressure 
scaling appears similar. Extending such an analysis to 2D systems is a promising 
direction for future work.

\jp{The two-step jamming transitions observed in Figs.~\ref{fig1} (a)–(b) resemble the 
emergence of multiple arrested states in thermal systems, such as those found in binary 
hard-sphere mixtures with varying composition \cite{voigtmann2011multiple}. While the 
glass-glass transition reported in monodisperse colloids with short-range attractions 
\cite{sperl2004dynamics} is driven by changes in interaction potential rather than 
composition, it exemplifies how distinct dynamical arrest mechanisms can coexist within 
a single system. In our case, the jamming process proceeds via two distinct stages: one 
dominated by the mechanical arrest of large particles, and a second involving the percolation 
of small particles into the jammed structure of large ones. This sequential arrest mirrors, in a 
geometric (athermal) context, the concept of multiple glassy states governed by different 
physical constraints. These analogies underscore the broader relevance of composition- and 
interaction-driven arrest, potentially informing unifying theoretical frameworks for both 
glass and jamming transitions \cite{charbonneau2017Glass, hexner2018Two}.}

\jp{The second jamming transition marks a qualitative shift in the packing structure and 
mechanical response of bidisperse granular materials. While the first transition is governed 
by the excluded area of large particles, the second arises from the saturation of voids by small 
particles—driven by their own excluded area—signaling a reorganization in space-filling 
and the onset of mechanical stability. The second transition reveals that jamming in 
mixtures of differently sized particles does not occur through a single, unified process. 
Instead, it can involve multiple, distinct transitions, each associated with a specific 
particle size and contributing uniquely to the emergence of rigidity through excluded area 
effects and spatial organization.}

We thank T. Kranz and Th. Voigtmann for proofreading, 
fruitful discussions, and providing constructive criticism about 
the results and the paper. This work was supported by the German 
Academic Exchange Service (DAAD) under grant n$^{o}$ 57424730.


\bibliography{Ref}
\bibliographystyle{apsrev4-2}

\end{document}


\begin{center}{\LARGE Supplementary Material: 
\\
\bigskip
Additional jamming transition in 2D bidisperse granular packings}
\linebreak[1]

Juan C. Petit\footnote{Juan.Petit@dlr.de}
and Matthias Sperl\\ 
Institut f\"{u}r Materialphysik im Weltraum, Deutsches Zentrum f\"{u}r Luft- und Raumfahrt (DLR), 51170 K\"{o}ln, Germany\\
Institut f\"{u}r Theoretische Physik, Universit\"{a}t zu K\"{o}ln, 50937 K\"{o}ln, Germany

\end{center}

\section{Technique and procedure of the simulation}
\label{SecI}

The open-source code {\tt www.MercuryDPM.org} is used for 2D Discrete Element
Method (DEM) simulations of frictionless soft-disk jammed packings without 
gravity \cite{cundall1979discrete, weinhart2020fast, petit2022bulk}. 
Newton's equations for each particle are solved numerically to predict its motion in time. 
$N = 5000$ particles are used to create bidisperse packings, where a number of large, 
$N_{\mathrm L}$, and small, $N_{\mathrm S}$, particles with radius 
$r_{\mathrm L}$ and $r_{\mathrm S}$ are considered. 
The size ratio is varied as $\delta = r_{\mathrm{S}}/r_{\mathrm{L}} \in [0.1, 0.9]$, 
while the concentration of small particles spans $X_{\mathrm{S}} = 
N_{\mathrm S} \delta^{2} / (N_{\mathrm L} + N_{\mathrm S}\delta^{2})  \in [0.05, 1]$.

We use the nonlinear Hertzian spring dashpot model based on the normal contact force 
between perfectly elastic particles, and it is defined by  
$\mathbf{f}^n_{ij}  = (k_n \alpha^{n}_{ij} 
+ \gamma_n \dot{\alpha}^{n}_{ij}) \hat{\bf n}$, see Ref.~\cite{weinhart2020fast}, where 
$k_n = (4/3) E^{*}_{ij} \sqrt{R^{*}_{ij}\alpha^{n}_{ij}}$ 
is the stiffness that depends on the effective Young's modulus 
$E^{*}_{ij} = E_{i}E_{j}/[(1-\nu^{2}_{i})E_{j} + (1-\nu^{2}_{j})E_{i}]$, with $E$ the Young 
modulus and $\nu$ the poisson ratio. The effective radius is defined by 
$R^{*}_{ij} = R_{i}R_{j}/(R_{i} + R_{j})$ and $\alpha^{n}_{ij}$ is the mean 
overlap between particle $i$ and $j$. The dissipative term considers the relative normal velocity $\dot{\alpha}_{n}^{c}$
and the viscoelastic normal dissipation coefficient $\gamma_n = \psi\sqrt{m^{*}_{ij} k_{n}}$, 
where $\psi = (\sqrt{5}\,{\rm ln}\,e)/\sqrt{{\rm ln}^{2}e + \pi^{2}}$ is the dissipation 
coefficient calculated by the restitution coefficient, $e$, introduced by hand.
$m^{*}_{ij} = m_{i}m_{j}/(m_{i} + m_{j})$ is the effective mass. 
Both small and large particles share the same material properties: 
density $\rho = 2500 \, [\mathrm{kg/m^3}]$, normal stiffness $k_n = 10^5 \, [\mathrm{kg/s^2}]$, 
and coefficient of restitution $e = 0.9$. These parameters are essential inputs 
for initializing simulations in MercuryDPM.

\begin{figure}[t]
\centering \includegraphics[scale=0.33]{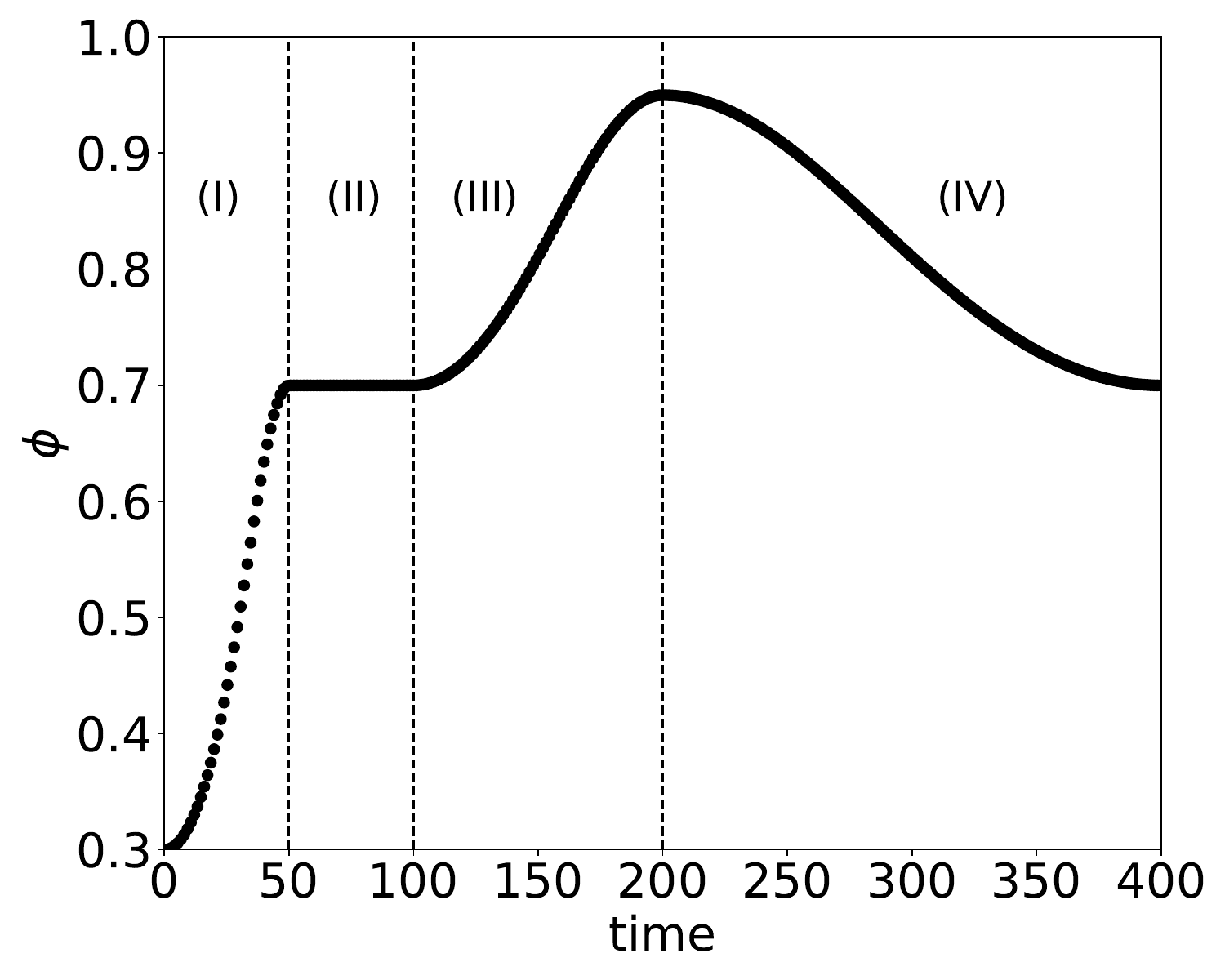}
\caption{Protocol of the simulation. (I) An initial compression is applied until $\phi_{0} =0.70$. (II)
A relaxation process is applied at a constant packing fraction to remove any kinetic energy. (III) 
The compression is then carried out after relaxation until $\phi_{\mathrm{max}}=0.95$. (IV) A decompression 
continues slower than the compression path until $\phi_{0}$ is reached again. }
\label{protocol}
\end{figure}

Each bidisperse packing, defined by $(\delta, X_{\mathrm{S}})$, is created and compressed/decompressed 
using a standardized protocol \cite{goncu2010constitutive, petit2022bulk}. The initial configuration
consists of disk particles with radii $r_{\mathrm{L}}$ and $r_{\mathrm{S}}$, 
uniformly distributed and randomly placed in a 2D square box without gravity. Particles 
are initialized at a packing fraction $\phi_{\mathrm{ini}} = 0.3$ with random velocities uniformly sampled from $[-10^{-2}, 10^{-2}]$. Large initial overlaps may cause a peak in kinetic energy, which is dissipated by the 
collisional damping, helping to randomize particle positions. This 
randomization process ensures that variations in initial conditions do not affect the 
jamming density. The granular gas is then isotropically compressed to reach an initial 
configuration with a target packing fraction $\phi_0 = 0.7$, slightly below the jamming density, 
see Fig.~\ref{protocol} (I). Afterward, the system undergoes relaxation at $\phi = \phi_0$, see (II). 
In this case, particles lose kinetic energy through inelastic collisions and background damping, allowing the system to naturally relax into a low-energy, disordered gas-like state prior to isotropic compression. This approach mimics the physical relaxation processes observed in granular materials. Then, a smooth isotropic compression (or decompression), see (III) and (IV), is 
applied up to $\phi = \phi_{\mathrm{max}} = 0.95$ (or back to $\phi = \phi_0$) at strain 
rate $\dot{\epsilon}\sim 6\times10^{-4}$ ($\dot{\epsilon}\sim 3\times10^{-4}$) 
by simultaneously moving all periodic boundaries inward (or outward).

The motion of the periodic boundaries is modeled using a cosine function. For example, the 
vertical wall position (or height) is given by 
\begin{equation}
z(t) = z_{f} + \frac{z_{0}-z_{f}}{2}(1 + \mathrm{cos}(2\pi\,(t-t_0)/T)),
\label{ecu0} 
\end{equation}  
where the strain is defined as $\varepsilon_{zz}(t) = 1 - z(t)/z_{0}$, with $z_{0}$ and $z_{f}$ 
the initial and maximum vertical dimensions of the box at zero and maximum strain, 
respectively. The total simulation time is $T = 400 \, [\mathrm{s}]$. 
During the compression and decompression paths, the strain rate is given by 
$\dot{\varepsilon}_v = \dot{\varepsilon}_{xx} 
+ \dot{\varepsilon}_{zz} \propto (1 - z_f/z_0)\, \mathrm{sin}(2\pi (t - t_0)/T)/T.$
The time-offset,  $t_0$, ensures a smooth start-up and finish of the boundary motion
via a co-sinusoidal function, avoiding shocks and inertia effects. Alternative testing 
methods could be used \cite{donev2006binary, o2003jamming, chaudhuri2010jamming}, but 
they would yield similar results, as the deformation occurs slowly—a regime where our  
procedure aligns with energy minimization, as shown in Ref.~\cite{krijgsman2016simulating}.
At the conclusion of the simulation protocol, the jamming density of each bidisperse 
packing is determined from the decompression branch. This approach is preferred as 
the decompression branch is less sensitive to deformation rates, as noted 
by Ref.~\cite{goncu2010constitutive}. Consequently, the decompression phase is 
intentionally longer (or slower) than the compression phase to get a precise estimate 
of $\phi_J$, as illustrated in Fig.~\ref{protocol} (IV).


\section{Fraction of large ($n_{\mathrm{L}}$) and small particles ($n_{\mathrm{S}}$) with $\phi$}
\label{SecII}

\begin{figure}[t]
\centering \includegraphics[scale=0.33]{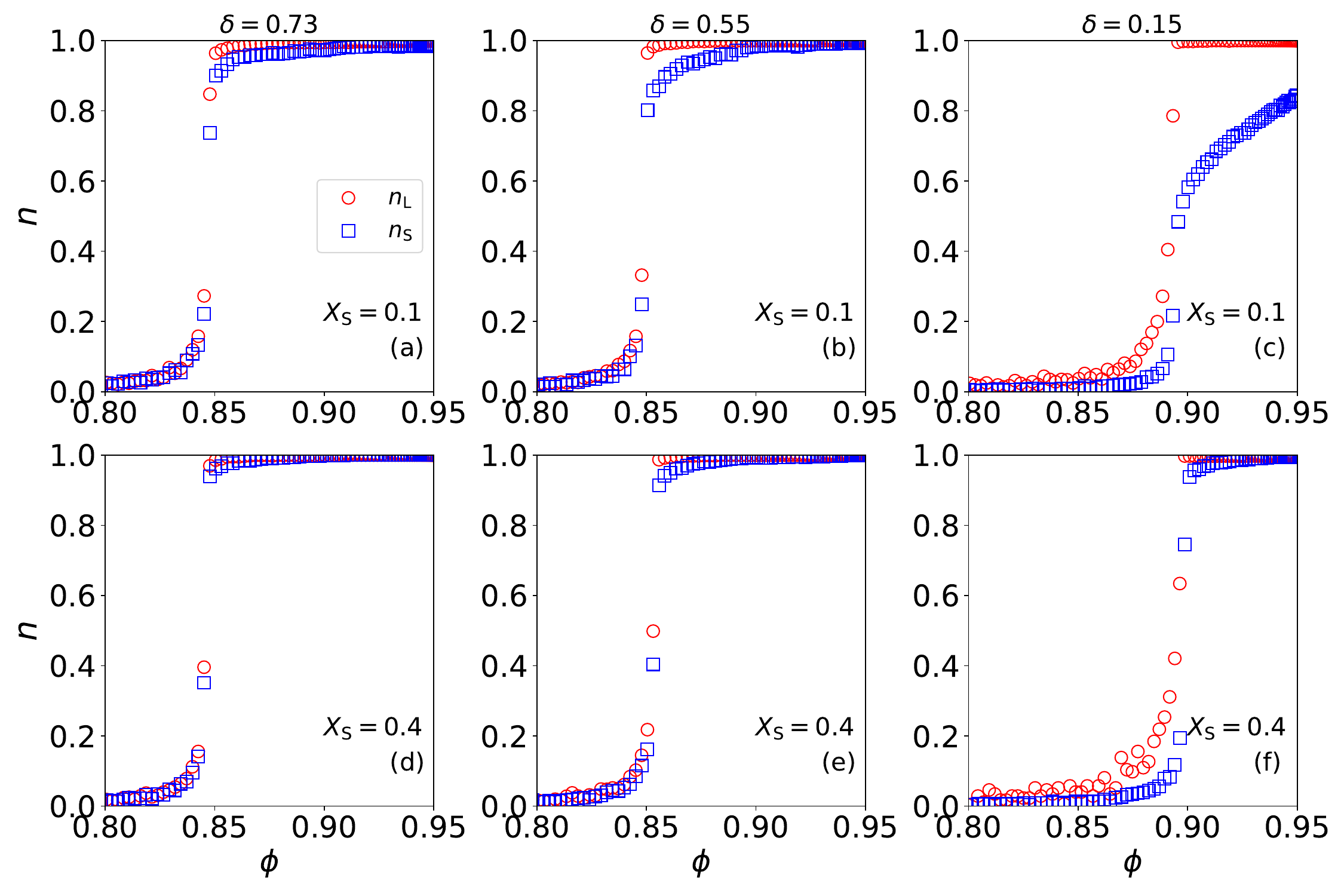}
\caption{Fraction of large, $n_{\mathrm L}$, and small, $n_{\mathrm S}$, particles contributing 
to the jammed structure as a function of the packing fraction $\phi$ for three $\delta \in [0.15,0.55, 0.73]$ 
values at $X_{\mathrm S} = 0.1$ and $X_{\mathrm S} = 0.4$. A slightly decoupling in the 
jamming density of the packing is observed for $\delta = 0.15$ and $X_{\mathrm S} = 0.1$.}
\label{fracPart}
\end{figure}

The fractions of large and small particles contributing to the jammed structure 
are quantified as $n_{\mathrm{L}} = N^{c}_{\mathrm{L}} / N$ and 
$n_{\mathrm{S}} = N^{c}_{\mathrm{S}} / N$, respectively, where $N^{c}_{\mathrm{L,S}}$ 
denotes the number of large and small particles in contact, and 
$N = N_{\mathrm{L}} + N_{\mathrm{S}}$ is the total number of particles. Jumps 
in $n_{\mathrm{L}}$ and $n_{\mathrm{S}}$ at specific $\delta$ values are shown 
in Fig.~\ref{fracPart} for $X_{\mathrm{S}} = 0.1$ and $X_{\mathrm{S}} = 0.4$.
For $\delta = 0.73$, the particles are similar in size, resulting in both large 
and small particles jamming simultaneously at the same density, irrespective of 
$X_{\mathrm{S}}$,  see Fig.~\ref{fracPart} (a, d). For intermediate size ratios, 
$\delta = 0.55$, both particle types still contribute to the jammed structure with 
minimal variation in $\phi_J$ as $X_{\mathrm{S}}$ changes, see Fig.~\ref{fracPart} (b, e). 
At $\delta = 0.15$, the size asymmetry allows small particles to fit into the voids 
of the triangular lattice of large particles, leading to a slightly different behavior 
compared to large particles. This is clearly seen in Fig.~\ref{fracPart} (c), where 
large particles jam first at low $\phi$, indicated by the jump in $n_{\mathrm{L}}$, 
followed by small particles, indicated by a jump in $n_{\mathrm{S}}$ at slightly 
higher $\phi$. These two features represent the first and second jamming transition
of the system. This behavior contrasts sharply with that shown in Fig.~\ref{fracPart} (f), 
where both $n_{\mathrm{L}}$ and $n_{\mathrm{S}}$ exhibit simultaneous jumps due to 
the higher $X_{\mathrm{S}}$.

Note for high $\delta$, $\phi_J$ is extracted at the packing fraction, $\phi$, where 
$n_{\mathrm{L}}$ or $n_{\mathrm{S}}$ jump (both jumps are equivalent). The same is true 
for low $\delta$ and high $X_{\mathrm{S}}$. However, at low $\delta$ and low $X_{\mathrm{S}}$, 
$\phi_J$ for large particles is similarly straightforward to identify, but for small particles, 
$\phi_J$ is defined when $n_{\mathrm{S}} \sim 0.6$ is reached. This occurs because the contribution 
of small particles to the jammed structure becomes substantial enough to decouple from the 
jamming of large particles. For $n_{\mathrm{S}} < 0.6$, 
the number of small particles contributing to the jammed structure is not enough to 
drive the system to an additional transition line. All extracted $\phi_J$ values are 
presented in Fig.~1 (a)-(b) of the paper, illustrating the distinct behaviors across 
the size ratios.


\section{Average overlap $\langle \alpha \rangle$ at first and second transitions}
\label{SecIII}

\begin{figure}[b]
\centering \includegraphics[scale=0.31]{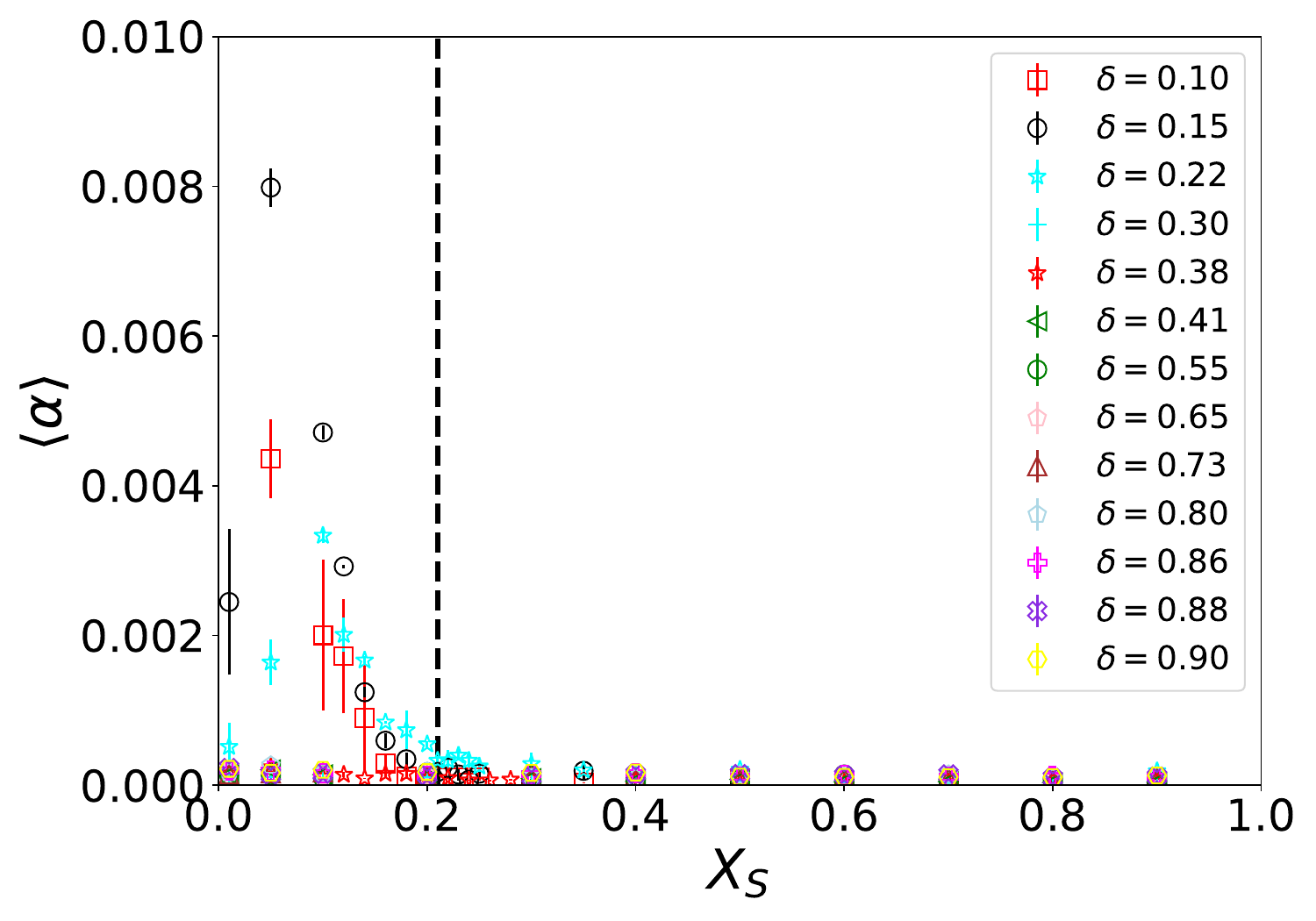}
\caption{Average overlap $\langle \alpha \rangle$ as a function of $X_{\mathrm{S}}$ 
for the $\delta$ values examined in this study. The dashed black line indicates the 
emerging point at $X_{\mathrm{S}}^{*}(\delta^{*}) \approx 0.21$ with $\delta^{*} \approx 0.25$. 
The sudden rise in $\langle \alpha \rangle$ for $X_{\mathrm{S}} < X_{\mathrm{S}}^{*}$ 
corresponds to the onset of the second jamming transition. Error bars represent the 
standard deviation computed from three independent realizations.}
\label{aveOverlap}
\end{figure}

We quantify the dimensionless average overlap as 
$\langle \alpha \rangle = \big(\sum_{ij}^{N_{c}} \alpha_{ij}\big)/2r_{\mathrm{L}}N_{c}$, 
where $\alpha_{ij}$ is the overlap between particles $i$ and $j$, $N_{c}$ is the total 
number of contacts, and $r_{\mathrm{L}}$ is the radius of the large particles. 
Fig.~\ref{aveOverlap} illustrates $\langle \alpha \rangle$ for each bidisperse 
packing at the first and second jamming transitions. Above the emerging point, 
represented by the dashed black line at $X_{\mathrm{S}}^{*}(\delta^{*}) \approx 0.21$, 
$\langle \alpha \rangle \sim 10^{-4}$, corresponding to the overlaps observed at 
the first jamming transition. In contrast, for $X_{\mathrm{S}} < X_{\mathrm{S}}^{*}$, 
where the second transition occurs, $\langle \alpha \rangle$ increases due to the 
incorporation of small particles and further penetration between large particles. 
Despite this increase, the overlaps along the second transition remain below $1\%$, 
providing strong evidence for the validity of the second jamming transition even after further 
compression.


\section{Furnas model in 2D}
\label{SecIV}

The jamming densities observed in Fig.~2 (a) in the paper are compared to 
an asymptotic model introduced by Furnas in Ref.~\cite{furnas1931grading},
suggesting that if the size ratio of the particles are extreme ($\delta \to 0$), 
$\phi_J$ can decouple in two limits sharing a common point at $X^{*}_{\mathrm S}$. 
One limit considers an approximation where large particles dominate the jammed 
structure while small particles are not taken into account since the number of 
them are not enough to play a role ($0 \leq X_{\mathrm S} < \hat{X}^{*}_{\mathrm S}$). 
In the second limit, both large and small particles participate in the jammed structure 
($0 \leq X_{\mathrm S} \leq 1$). In this case, the number of small particles are high 
enough to drive few large particles to the jammed state. Both limits are written as 

\begin{equation}
\lim_{\delta \to 0} \phi_{J}(X_{\mathrm S}) =
  \begin{cases}
    \frac{\phi_{J}^{\rm mono}}{(1 - X_{\mathrm S})} \,\, & \text{if } 0 \leq X_{\mathrm S} < X^{*}_{\mathrm S},\\
     \frac{\phi_{J}^{\rm mono}}{[\phi_{J}^{\rm mono} + (1 - \phi_{J}^{\rm mono})\,X_{\mathrm S}]}  &  \text{if } 0 \leq X_{\mathrm S} \leq 1,
  \end{cases}
\label{ecu1}
\end{equation}

\begin{figure}[t]
\centering \includegraphics[scale=0.35]{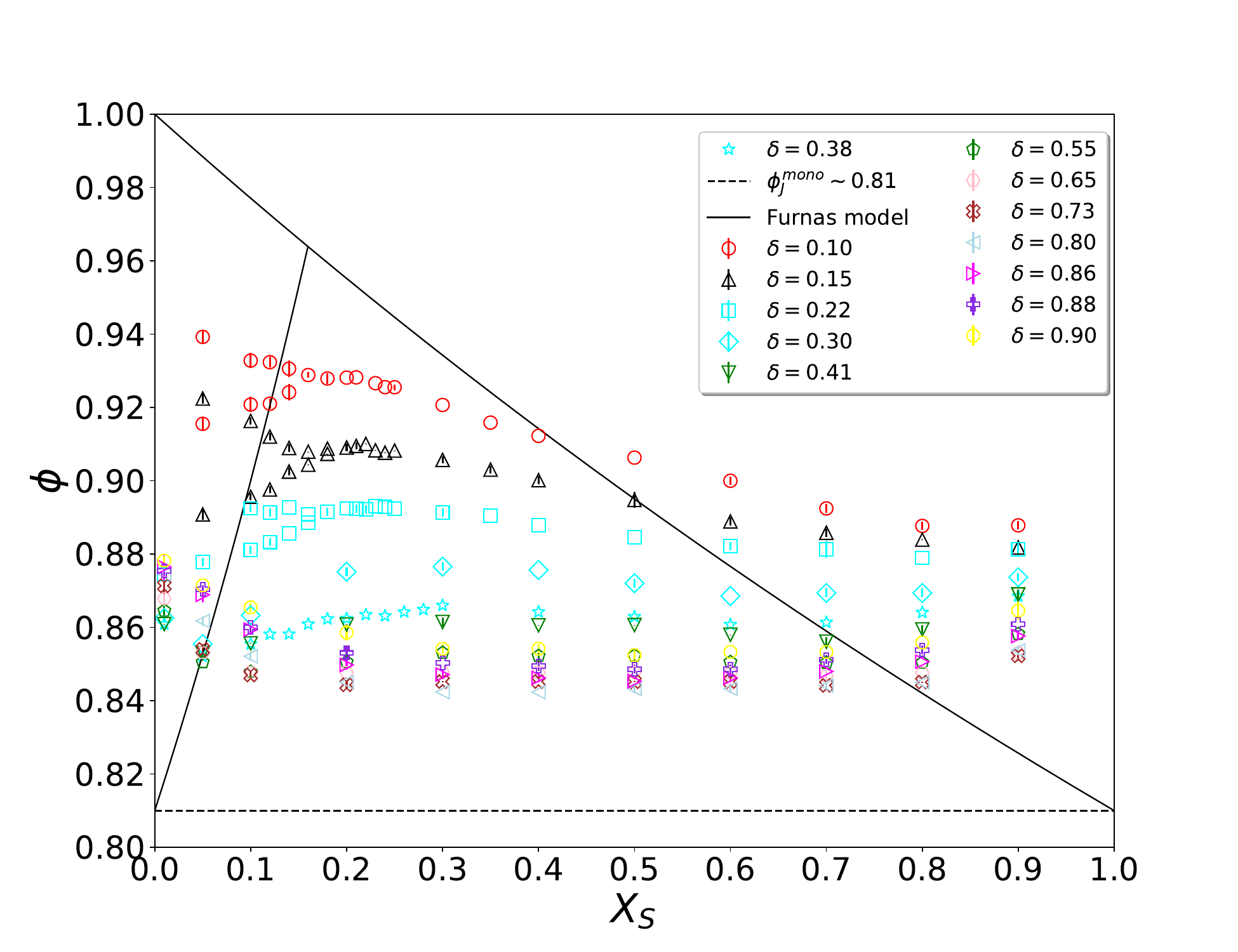}
\caption{Jamming densities, $\phi_J$, as a function of the concentration of small particles,
$X_{\mathrm{S}}$, for the range of size ratio, $\delta$, used in the paper. The black solid line represents the Furnas model for 2D packings given in Eq.~\eqref{ecu1}. Error bars represent the 
standard deviation computed from three independent realizations, with values remaining below $10^{-2}$.}
\label{furnas2d}
\end{figure}

\medskip

\noindent and are displayed in Fig.~\ref{furnas2d} with $\phi_{J}^{\rm mono} \approx 0.81$. We observed
that the Furnas model does not follow nor the first either the second jamming transitions as shown 
previously in 3D results \cite{petit2020additional, petit2022bulk}. Although, it shows a maximum $X^{*}_{\mathrm S}=(1-\phi_{J}^{\rm mono})/(2-\phi_{J}^{\rm mono}) \approx 0.16$ near to $X^{*}_{\mathrm S} \approx 0.21$ obtained for $\delta = 0.1$. This discrepancy highlights the need to either revise the Furnas model or develop an alternative framework better suited to two-dimensional systems.

\section{Fraction of small particles at the second transiton ($n^{2}_{\mathrm{S}}$)}
\label{SecIV}

\begin{figure}[t]
\centering \includegraphics[scale=0.29]{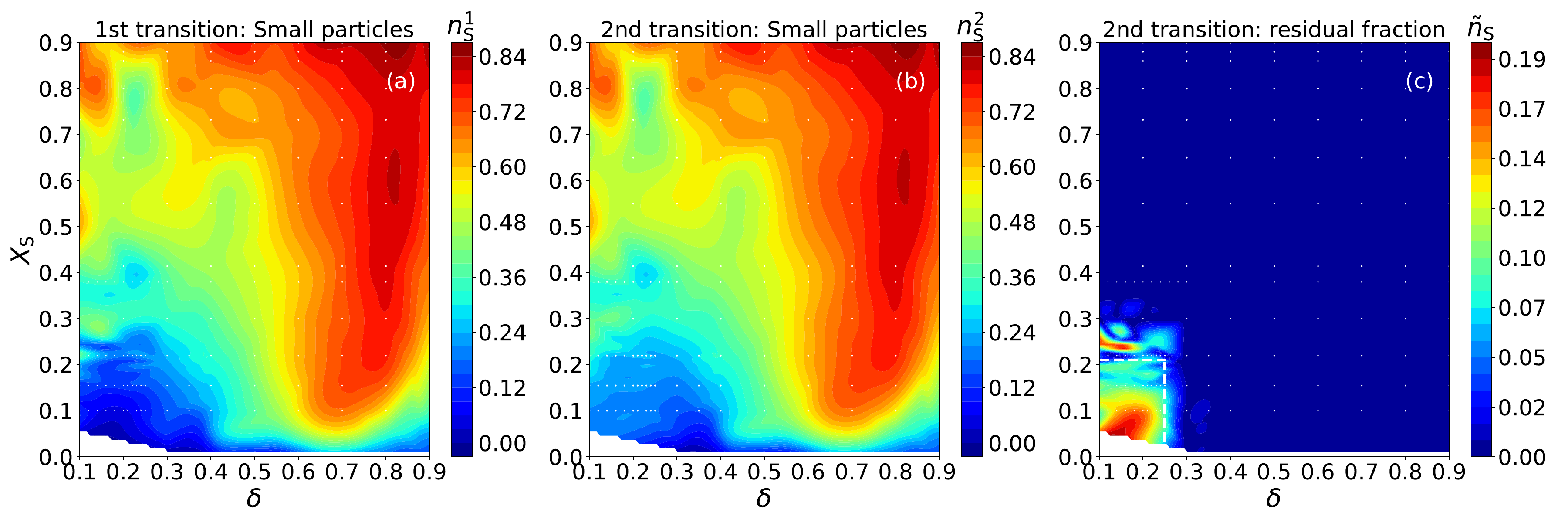}
\caption{Color maps in the fraction of small particles with $Z \geq 4$ at (a) the first jamming transition, 
(b) second jamming transition, and (c) residual fraction at the second transition quantified 
by $\tilde{n}_{\mathrm{S}} = n^{2}_{\mathrm{S}} - n^{1}_{\mathrm{S}}$. The white dashed lines 
a low $\delta$ and low $X_{\mathrm{S}}$ contain the second jamming transition. }
\label{residualFract}
\end{figure}

Fig.~\ref{residualFract} presents color maps of the fraction of small particles with $Z \geq 4$ at (a) the first 
jamming transition and (b) the second jamming transition. Both maps exhibit similar patterns, with 
subtle differences observed at low $\delta$ and low $X_{\mathrm{S}}$. To quantify the regions 
where the number of small particles changes between the first and second transitions, we calculate 
the difference $\tilde{n}_{\mathrm{S}} = n^{2}_{\mathrm{S}} - n^{1}_{\mathrm{S}}$, shown in 
Fig.~\ref{residualFract} (c). A distinct region at low $\delta$ and low $X_{\mathrm{S}}$ 
emerges, indicating that this is the only area where the small particle fraction varies between 
transitions, while the rest remains unchanged. This region is directly associated with the second 
transition, with the emerging point identified at $X^{*}_{\mathrm{S}}(\delta^{*}) \approx 0.21$ 
and $\delta^{*} \approx 0.25$, see the white dashed lines. This boundary separates areas with a substantial number of small particles from regions with low 
$\tilde{n}_{\mathrm{S}}$. For large particles, the behavior is 
straightforward: all large particles jam during the first transition, displaying the same pattern 
at the second transition. The fraction of large particles at the first transition is shown in 
Fig.~3 (a) of the paper.

\bibliographystyle{unsrt}
\bibliography{RefSuplement}